\documentclass[12pt,a4paper,final]{iopart}

\usepackage{iopams}  
\usepackage{graphicx}
\usepackage{epstopdf}
\usepackage[breaklinks=true,colorlinks=true,linkcolor=blue,urlcolor=blue,citecolor=blue]{hyperref}
\begin{document}

\title[]{Reduction of residual stress in AlN thin films synthesized by magnetron sputtering technique}
\author{Padmalochan Panda$^a$, R. Ramaseshan$^{a*}$, N. Ravi$^b$, G. Mangamma$^a$, Feby Jose$^a$, S. Dash$^a$, K. Suzuki$^c$ and H. Suematsu$^c$}
\address{$^a$Surface and Nanoscience Division, Materials Science Group, Indira Gandhi Centre for Atomic Research, Kalpakkam - 603102, India.}
\address{$^b$Center for Engineered Coatings, ARCI, Hyderabad - 500005, India.}
\address{$^c$Extreme Energy Density Research Institute, Nagaoka University of Technology, Nagaoka, Japan.}
\ead{seshan@igcar.gov.in}
\begin{abstract}
We report the reduction in residual stress of AlN thin films and also the crystal structure, surface morphology and nanomechanical properties of magnetron sputtered as a function of substrate temperature ($T_s$, 35 - 600 $^\circ$C). The residual stress of these films was calculated by $sin^{2}\psi$ technique and found that they are varying from tensile to compression with temperature ($T_s$). Evolution of crystalline growth of AlN films was studied by GIXRD and transmission electron microscopy (TEM) and a preferred a-axis orientation was observed at 400 $^\circ$C. The cross-sectional TEM micrograph and selected area electron diffraction (SAED) of this film exhibited a high degree of orientation as well as a columnar structure. Hardness $(H)$ measured by Nanoindentation technique on these films ranged between 12.8 - 19~GPa.
\end{abstract}

\submitto{\JPD}

\section{Introduction}

Aluminum nitride (AlN) is a versatile material with great technological advantages for semiconductor industry due to its wide band gap ($\thicksim$ 6.2 eV), high thermal conductivity (up to 320 W/mK) and small thermal-expansion coefficient \cite{Hadis,Pan}. It has been viewed as a highly promising material system for high-temperature (melting temperature $\thicksim$ 3000~$^\circ$C) optoelectronic and short wavelength light source/detector applications \cite{Barkad}. It is used as optical films due to its high refractive index ($\thicksim$ 2.0 at a wavelength of 520 nm) as well as good dielectric constant ($\thicksim$ 7 at a frequency of 300 kHz) \cite{Dimitrova}. Due to the similarity in crystal structure and lattice constant between AlN and GaN, the AlN film is also used as beneficial buffer layer for the growth of GaN films for optoelectronic devices \cite{Matt}. Highly efficient deep-ultraviolet light-emitting diode (UV-LED) is an expected future application where AlN thin films find an important role with emission wavelength of 210 nm (shortest wavelength ever observed from any semiconductor) using a p-i-n homojunction of a-plane oriented AlN. It is expected to get higher emission intensity from  a-axis AlN than c-axis AlN due to its an-isotropic emission pattern and its lower c/a ratio \cite{Taniyasu}. Inappropriately, the lattice mismatch between AlN thin film with the substrate induce threading dislocations due to stress, which are of the order of $10^{9}$ to $10^{11}$  $cm^{-2}$ in the epilayer \cite{Pei}. Due to the existence of threading dislocations, its external quantum efficiency (EQE) is very low and have influence on device performance not only in LED but also in high-power, high-frequency electronic devices and bulk acoustic wave resonator applications \cite{Shibo}. 

In the past decades, an extensive research work has been carried out on optimization of deposition by various coating techniques with deposition parameters in order to obtain the required reproducible quality thin films. Epitaxial growth of AlN has already been demonstrated on various monocrystalline substrates such as growth by chemical vapor deposition  on (0001) oriented AlN substrate, sandwich sublimation configuration to grow AlN epitaxy, molecular beam epitaxy on GaN templates and Si (111) and deposition by pulsed laser deposition  on molybdenum \cite{Bryan,Georgieva,Kuyoma,Okamoto}. In most techniques, the deposition temperatures are quite high, so a smooth surface morphology could not be obtained due to the degradation of substrate and thermal damage of AlN thin films during deposition \cite{Liu}. For SAW device applications, a highly oriented film and homogeneous composition with low surface roughness is expected, since large surface roughness may lead to increase in propagation loss especially when high frequency is reached \cite{Ping}. Reactive sputtering technique is promising under such circumstances when, low substrate temperature deposition (wide variety of substrate materials, compatible with current semiconductor device processes) and good surface finish as well as adhesion, are required \cite{Mirpuri}. It is reported in the literature that AlN films deposited at low target to substrate distance (TSD) and low sputtering pressure exhibited (002) preferred orientation whereas those synthesized at high TSD and high pressure exhibited (100) orientation \cite{Xu,Feby}. However, for applications, the fabrication of AlN thin film devices need a better understanding of the mechanical properties in addition to its electrical and optical performances, since the contact loading during processing or packaging can significantly degrade the performance of these devices. So a precise measurement of nano-mechanical characteristics of AlN films is required, where nanoindentation has been widely used to study the hardness, modulus and the deformation mechanisms \cite{Soh,Jose}.

So, in this work, we have primarily concerned with the residual stress in AlN films by varying substrate temperature and have reported the evolution of orientation, surface roughness and mechanical properties.

\section{Experimental}

AlN thin films were deposited on silicon (100) substrates by DC reactive magnetron sputtering technique, using MECA 2000 (France) from  pure aluminum target (4N) of 50 mm diameter in a high purity argon (4N) and nitrogen (4N) gas mixture. P-type Si~(100) substrates were cleaned by a standard two-step RCA process, boiled in a solution of H$_2$O : NH$_4$OH : H$_2$O$_2$ (5:1:1). Then, they were etched in 2$\%$ HF-H$_2$O solution for 2~min to remove the native oxide layers on the surfaces. Before deposition, the sputtering chamber was evacuated to $1 \times 10^{-6}$ mbar. First, in Ar atmosphere a mono layer of Al was deposited on the substrate for few seconds to improve the adhesion between the substrate and deposited AlN thin film. These films were grown at different substrate temperatures and along with constant ratio of sputtering (Ar) to reactive gas (N$_2$), target to substrate distance (TSD) and deposition pressure. The deposition parameters in this study are listed in Table \ref{tab:1}. A surface profilometer (M/s. Dektak 6M, Veeco, USA) was used to measure the thickness of films, was around 1 $\mu m$.

\begin{table}[h] 
\centering
\begin{tabular}{l l}
\hline\hline
Sputtering parameters  &\multicolumn{1}{c}{Values} \\[0.5ex]
\hline
Target ~&~ Al (4N pure)\\
Substrate ~&~ Si (100)\\
Deposition pressure (mbar) ~&~ $5 \times 10 ^{-3}$ \\
TSD (cm) ~&~ 14 \\
Ar : N$_2$ (sccm) ~&~ 4 : 1 \\
DC power (W) ~&~ 200 \\
Substrate temperature ($T_s$) ($^\circ$C) ~&~ 35 to 600\\[1ex] 
\hline\hline
\end{tabular}
\caption{\label{tab:1}Deposition parameters of AlN films at different growth temperatures.}
\end{table}

Identification of crystalline phases was carried out by Bragg-Brantano geometry and Grazing Incidence X-ray Diffraction (GIXRD) technique (M/s. Bruker D8, Germany) at a grazing angle of 0.5$^\circ$C. Bragg - Brentano (Cu - $K_\alpha$) geometry was used to measure the residual stress of these films. The residual stress measurements of these films were carried out using $sin^{2}\psi$ technique with angle inclination ($\psi$) varied from 0$^\circ$ to $\pm$ 90$^\circ$ with step size of 4$^\circ$. Cross section samples for transmission electron microscopy have been prepared using a focused ion beam apparatus (JIB-4500, JEOL, Japan). Cross sectional microstructural studies were performed using a transmission electron microscopy (M/s JEM-2100F, JEOL, Japan ) operating at 200 kV. The surface morphology of these films was characterized by an atomic force microscope (NTEGRA Prima, M/s. NT-MDT, Russia) with a contact cantilever single-crystal silicon tip of size 10 nm. A TriboIndenter (TI 950, M/s. HYSITRON, USA) equipped with a three sided pyramidal diamond berkovich tip was used to study the nanomechanical properties of these films with the peak load of 4000 $\mu$N, with a loading/unloading time of 5 sec and 2 sec holding time. The hardness $(H)$ and modulus $(E)$ values of these films were calculated using Oliver and Pharr model \cite{Feby}.

\section{Results and Discussion}
\subsection{XRD}
\begin{figure}[h]
\centering
\includegraphics[width=0.60\textwidth]{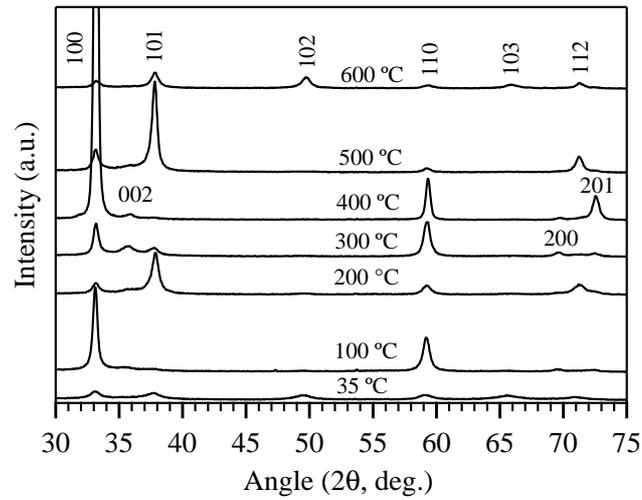}
\caption{GIXRD profiles of AlN thin films with different substrate temperatures.}
\label{fig:1}
\end{figure}

\begin{figure}[h]
\centering
\includegraphics[width=0.60\textwidth]{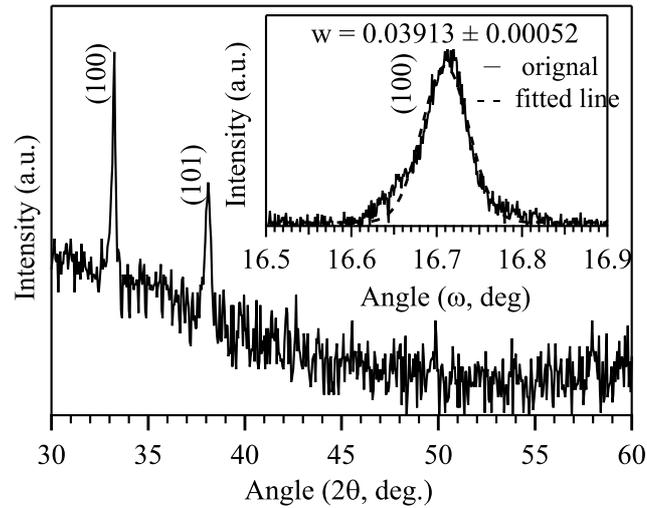}
\caption{Displays the XRD pattern of 400 $^\circ$C substrate temperature AlN thin film in B-B geometry and a rocking curve for 100 planes is shown in the inset with FWHM.}
\label{fig:2}
\end{figure}

The GIXRD profiles of AlN thin films sputtered on Si (100) with different substrate temperatures varying from 35 to 600 $^\circ$C by DC magnetron sputtering are shown in Fig.\ref{fig:1}. The GIXRD patterns of these films show poly-crystalline hexagonal wurtzite structure and agree with the JCPDS data 25-1133. At 35 $^\circ$C, a poly-crystalline nature was observed, however with the increase in temperature from 100 to 500~$^\circ$C, the crystallinity has increased along with a preferential orientation towards either (100) or (101). At 100 and 400 $^\circ$C, the films exhibited a texture towards a-axis (100), whereas 200 and 500~$^\circ$C films exhibited (101). Among all, at 400 $^\circ$C substrate temperature exhibited a clear preferential orientation towards (100). A symmetrical diffraction study (B-B geometry)has been carried out to verify the growth direction, shown in Fig.\ref{fig:2}, also exhibited a preferential orientation at (100). The inset of Fig.\ref{fig:2}, a rocking curve of (100) exhibited a FWHM of 0.01413$\pm$0.00052 at the angle 33.2. However this is larger than the one reported earlier when the target to distance was increased \cite{Feby}. This signifies the 100 planes are parallel to the substrate. When a film is irradiated in asymmetrical diffraction configuration (GIXRD), such that the angle between certain planes of the oriented crystals and X-ray beam is equal to Bragg angle of that family of planes, a high intensity peak should be observed and it is called texture. On the other hand, if there is no high intensity measured in symmetrical diffraction (B-B geometry) for such orientation, still there is chance to conclude that there is a preferential orientation in the film provided other orientations are absent. Such crystals have non-random orientations but not parallel to the surface of the substrate to be observed as epitaxy. However, epitaxial growth is crystals grown with similar orientation in the three directions of space. In general, texture growth is nothing but the crystals inside a polycrystalline film grown in non-random crystallographic directions.

\begin{figure}[h]
\centering
\includegraphics[width=0.60\textwidth]{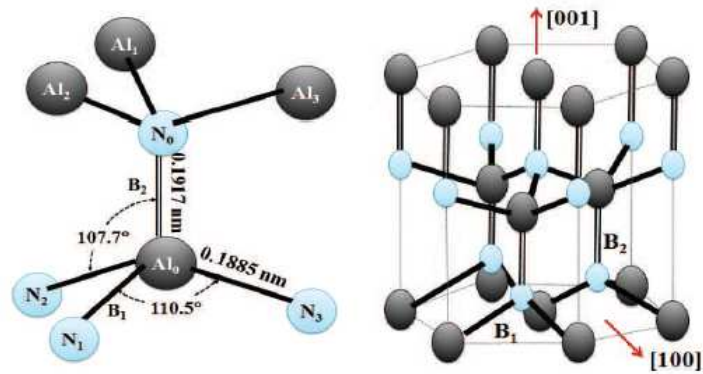}
\caption{(Color online) Crystal structure of wurtzite AlN.}
\label{fig:3}
\end{figure}

To understand the growth of AlN thin films with temperature, it is important to know the bonding nature and  strength with its orientation. Wurtzite structure AlN is shown in Fig.\ref{fig:3}, each Al atom is linked by four N atoms forming a distorted tetrahedron with three Al-N$_{(i)}$ (i = 1, 2, 3) bonds, namely B$_1$, the bond length is 0.1885 nm and one Al - N$_0$ bond in the c - axis direction, namely B$_2$, the bond length is 0.1917 nm. The bond B$_2$ is more ionic character  due to the coupling of Al empty orbit and the N full orbit, whereas B$_1$ is more covalent in nature due to $sp^{3}$ - hybridization between semi - full orbits of Al and N atoms \cite{Xu,Medjani}. The (100) plane is composed of the bonds B$_1$ which is more stronger as compared to B$_2$ bonds. So at such a low pressure and increasing temperature, B$_2$ bond can be easily dissociated which led the increasing rate of vertical growth of the (100) planes. At 400 $^\circ$C substrate temperature, the film is highly oriented along a-axis due to the (100) crystallites overgrowth the all other crystallites. But above 400 $^\circ$C, different orientations are present, since higher temperature is more favorable for the dissociation and regrowth of both the bonds. Therefore, it can be understood that the optimum substrate temperature for depositing highly a-axis oriented AlN films is~400~$^\circ$C.

\begin{figure}[h]
\centering
\includegraphics[width=0.60\textwidth]{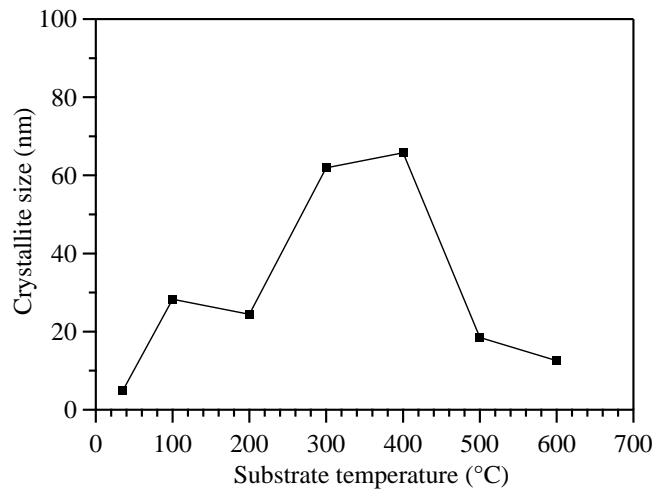}
\caption{The variation of crystallite size of AlN thin films with different substrate temperatures.}
\label{fig:4}
\end{figure}

Crystallite size has been calculated from GIXRD profiles using Williamson-Hall method, plotted against substrate temperature is shown in Fig.\ref{fig:4}. This increases with the increase of substrate temperature upto 400 $^\circ$C after which it started to decrease. This is due to the enhanced atom mobility at high temperatures, that this increases the ad-atoms surface diffusion length leading to a higher probability of ad-atoms migrate from the landing site to proper site. So that a better crystallinity and higher degree of orientation is seen up to 400 $^\circ$C. However, further increase in the substrate temperature the crystallite size decreased drastically. At higher temperatures over 400 $^\circ$C, excessive surface mobility of ad-atoms make thin film texture difficult due to stronger re-evaporation effect which results in reducing the crystallite size and intensity of the peaks \cite{Kuan}.

\subsection{Cross sectional TEM}
\begin{figure}[h]
\centering
\includegraphics[width=0.60\textwidth]{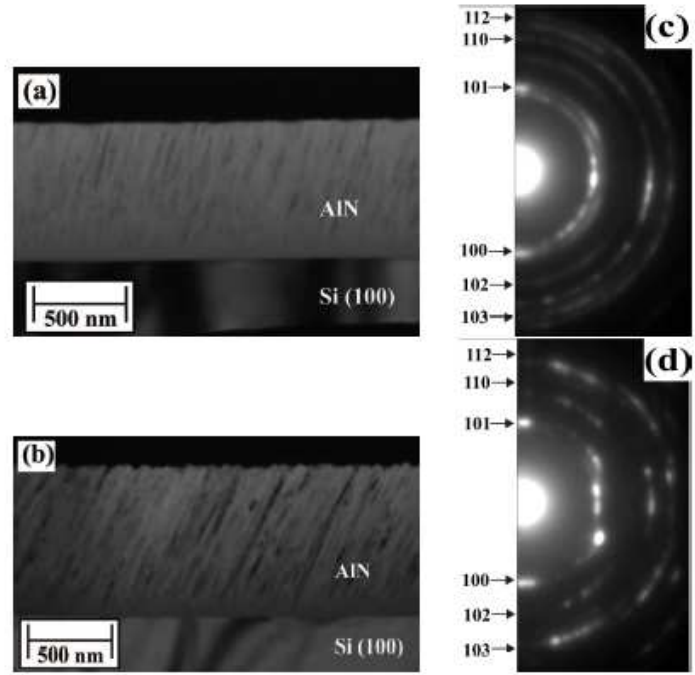}
\caption{The bright-field cross-sectional TEM image of AlN films deposited on Si at (a) 100 $^\circ$C, (b) 400 $^\circ$C substrate temperatures and corresponding SAED patterns, respectively.}
\label{fig:5}
\end{figure}

Figs.\ref{fig:5}(a) and (b) show bright-field TEM cross sectional micrographs of AlN film grown at 100 and 400 $^\circ$C substrate temperatures, respectively. The structure comprises of parallelly spaced columnar structure with an angle tilted by a few degrees to the substrate normal. In our sputtering system there are three guns placed focussing towards the substrate at an angle.  So the sputtered atom flux is arriving obliquely at an angle $\alpha$ and causes the inclination of the columns by $\beta$ where $\alpha$ and $\beta$ are related by the empirical tangent-rule : $\tan\beta$ = 1/2 tan$\alpha$ \cite{Mahieu}. The interface between the coating and the Si substrate is very sharp in both these films. A thin layer of Al deposited to improve the bonding between AlN and Si substrate is visible at the bottom of the columnar AlN films. A complete list of TEM micro-graphs and its respective SAED images are attached in the supplementary data (A). 

Figs.\ref{fig:5}(c) and (d) show the corresponding SAED patterns of those films discussed above deposited at 100 and 400 $^\circ$C substrate temperatures, respectively. The diffraction rings are indexed using the space group P$6_3$mc, lattice constants a=0.311 nm and c=0.497 nm (JCPDS file 25-1133). The one deposited at 100 $^\circ$C exhibited a polycrystalline micro-structure with mixed planes of (100), (101), (102), (110), (103) and (112). However, the one deposited at 400 $^\circ$C exhibited a preferential orientation along (100). This orientation has exhibited a pattern, since a polycrystalline random orientated planes exhibit a perfect circular ring. This pattern exhibited a well crystallized with a preferential orientation towards a-axis than  100 $^\circ$C grown film. Also,  SAED patterns of 100 and 400 $^\circ$C substrate temperature films are analogous to the GIXRD profiles shown in Fig.\ref{fig:1}. 

\subsection{Morphology and surface roughness by AFM}
\begin{figure}[h]
\centering
\includegraphics[width=0.60\textwidth]{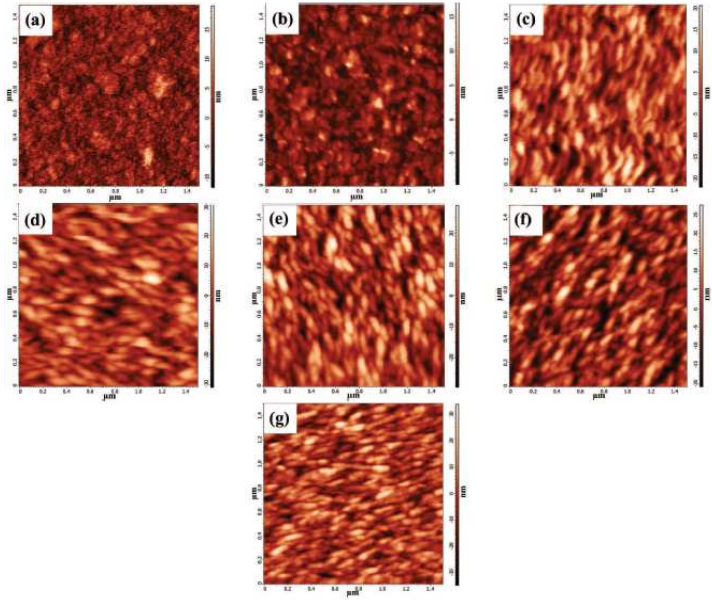}
\caption{(Color online)AFM images of AlN films on Si (100) with growth temperatures (a) 35 $^\circ$C, (b) 100 $^\circ$C, (c) 200 $^\circ$C, (d) 300 $^\circ$C, (e) 400 $^\circ$C, (f) 500 $^\circ$C, and (g) 600 $^\circ$C.}
\label{fig:6}
\end{figure}

Due to its excellent piezoelectric response, AlN thin films are highly sought after for in SAW devices. Surface acoustic waves cannot pass through the surface of a material if the surface roughness is greater than the wavelength \cite{Ping}. So its micro-structure and surface roughness have a crucial role on the performance of the devices. Fig.\ref{fig:6} consists of AFM images taken over 1.5 $\times 1.5~\mu m^{2}$ surface area showing the morphology of the AlN films deposited on Si at various substrate temperatures. In Fig.\ref{fig:6} (a and b), AlN films deposited at 35 and 100 $^\circ$C exhibit irregular grains with less surface density. However,  the increase in substrate temperature from 200 $^\circ$C to 600 $^\circ$C resulted a similar morphology with a pebble-like grain structure and dense surface.

\begin{figure}[h]
\centering
\includegraphics[width=0.60\textwidth]{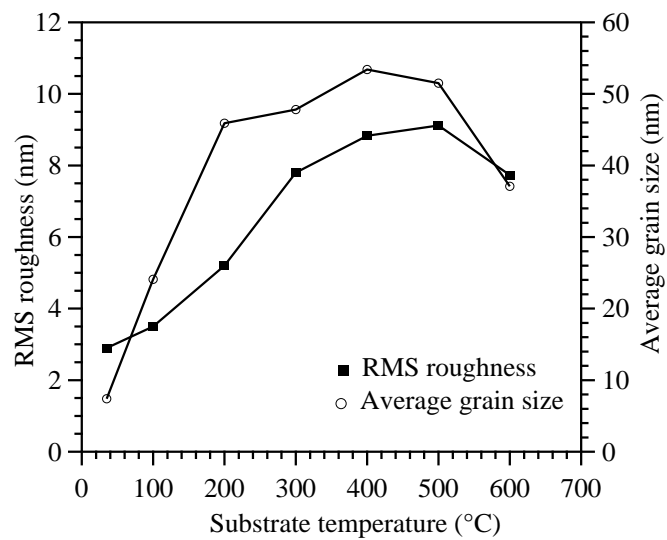}
\caption{The surface roughness RMS values and average grain sizes measured from the AFM images versus the substrate temperatures of AlN films.}
\label{fig:7}
\end{figure}

Fig.\ref{fig:7}, displays the root mean square (RMS) roughness and the mean diameter of grains against the substrate temperature. RMS roughness and average grain size are significantly affected by the increase in substrate temperature. Since the thermal energy is delivered to the surface, the energy and mobility of surface ad-atoms are enhanced and the atomic diffusion. This resulted in the formation of fewer large grains by agglomeration of adjoining atoms or crystals  and make surface coarser with the substrate temperature. This can be seen in the cross-sectional TEM image (Fig.\ref{fig:5}), the waviness is pronounced more in 400~$^\circ$C. But, beyond 500 $^\circ$C roughness decreases due to the decrease in grain size and crystallite size whichis  similar to Fig.\ref{fig:4}. Generally, the surface roughness of the films is expected to be less than 30 nm for surface acoustic wave devices. From this point of view, these film's surface roughness is adequate for the SAW and bulk acoustic wave (BAW) devices.
 
\subsection{Residual stress analysis of AlN films}
Residual stress plays an important role on thin films since it alters the physical properties like energy-band structure, external quantum efficiency of LED, piezoelectric constant and mechanical properties. This may cause micro-cracks or peeling from the substrate as well as the poor performance and altered lifetime of the coated component. Therefore, it is important to control the residual stress by controlling the deposition parameters to synthesize a mechanically stable films.

Measurement of residual stress (in-plane stress) has been carried out by a well-documented $sin^{2}\psi$ technique where crystallographic planes tilted at multiple angles ($\psi$), from the surface normal \cite{Polaki}. The lattice spacing depends on the strain in a elastically strained crystalline material. There is a correlation exists between the stress components in the film and variation in d-spacing with tilt angle ($\psi$). The slope of the linear relationship between d-spacing and $sin^{2}\psi$ is directly proportional to the stress component operative in the plane of the specimen along the direction of the diffraction vector according to the equation shown below 

\begin{equation}
\frac{d_{\psi}-d_0}{d_0} = \left[\left(\frac{1+\nu}{E}\right)_{\left(hkl\right)} \sigma_\varphi sin^{2}\psi \right] -  \left[{\left(\frac{\nu}{E}\right)}_{\left(hkl\right)} {\left(\sigma_1 + \sigma_2\right)}\right] 
\end{equation}

where E - Young's modulus, $\nu$ - Poisson's ratio, $\left(\frac{d_\psi - d_0}{d_0}\right)$ - micro-strain, $\left(\frac{1 + \nu}{E}\right), \left(\frac{-\nu}{E}\right)$ are X-ray elastic constants and $\sigma_1, \sigma_2$ are principal stresses. The above equation (1) can be reduced to:
 
\begin{equation}
d_\psi = \left[{\left(\frac{1 + \nu}{E}\right)}_{\left(hkl\right)} d_0\sigma_\varphi sin^{2}\psi\right] + d_0
\end{equation}

\begin{figure}[h]
\centering
\includegraphics[width=0.60\textwidth]{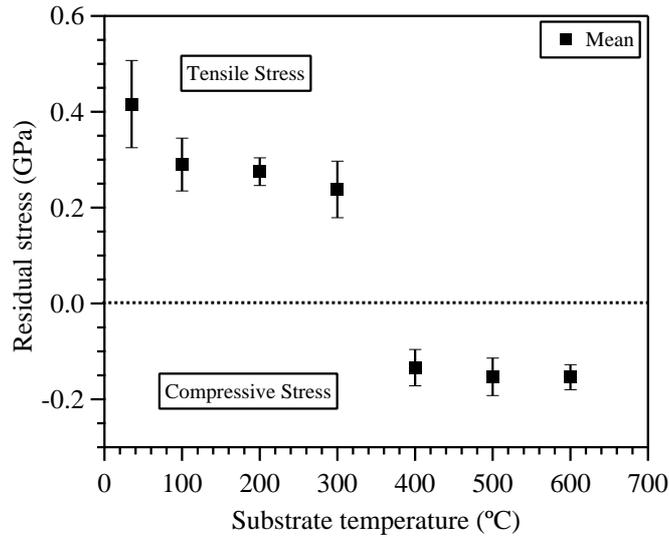}
\caption{Residual stress of AlN thin films as a function of substrate temperatures.}
\label{fig:8}
\end{figure}

The observed residual stress of all these films are plotted as a function of substrate temperature shown in Fig.\ref{fig:8}. A tendency from tensile to compression is observed with the increase in substrate temperature. The mean value of the residual stress has decreased from 0.42 GPa to 0.27 GPa, when the substrate temperature increased from 35 $^\circ$C to 300 $^\circ$C. However, above this temperature a transition has occurred in the internal stress, tensile to compressive mode. At 600 $^\circ$C, the residual stress has increased in the compression mode up to 0.15 GPa. 

To understand the nature of residual stress with the deposition parameters, the literature results are summarized in Table \ref{tab:2} and Table \ref{tab:3} for substrate temperature and power/pressure, respectively. Especially, only the transition from tensile to compressive stress regions and vice versa are picked up for the discussion. Among all, only Wang \emph{et al.} reported a similar transition (the bias voltage was kept at -40 V), whereas in our case the bias voltage was 0 V \cite{Wang}. It is clearly evident that increase in bias voltage will increase the rate of deposition till it reaches a critical value and after it decreases. Also, the internal stress is relatively independent of film thickness. 

\begin{table}[h] 
\centering
\begin{tabular}{| p{4.5cm}  | p{4cm}  | p{4.5cm}  | p{1cm} |} 
\hline\hline
RT to high temperature residual stress variation &  Range of shift ($T_s$ )& Deposition technique & Ref.\\
\hline
Tensile to compressive & 150 to 300 $^\circ$C & DC reactive sputtering & \cite{Wang}\\
\hline
Compressive to tensile & RT to 400 $^\circ$C & RF reactive sputtering & \cite{Medjani}\\
\hline
Tensile to compressive & 300 to 400 $^\circ$C & DC reactive sputtering & this work\\
\hline\hline
\end{tabular}
\caption{\label{tab:2}Residual stress of AlN films at different substrate temperatures.}
\end{table}

\begin{table}[h] 
\centering
\begin{tabular}{| p{5cm}  | p{4cm}  | p{4.5cm}  | p{0.5cm} |} 
\hline\hline
Residual stress variation with sputtering power and pressure &  Range of shift (Power/ Pressure) & Deposition technique & Ref.\\
\hline
Tensile to compressive & 6 - 6.5 kW & RF reactive sputtering & \cite{Knisely}\\
\hline
Compressive to Tensile & 3.99 - $5.32 \times 10 ^{-3}$ mbar  & pulsed DC sputtering & \cite{Dubois}\\
\hline
Compressive to tensile & 6.6 - $13.3 \times 10 ^{-3}$ mbar & RF reactive sputtering & \cite{Si}\\
\hline
Compressive to tensile & 6 - $8 \times 10 ^{-3}$ mbar & DC reactive sputtering & \cite{Liu}\\
\hline\hline
\end{tabular}
\caption{\label{tab:3}Residual stress of AlN films at different sputtering power and pressure.}
\end{table}

At low $T_s/T_m$ ($T_s$ = substrate temperature and $T_m$ = melting temperature of coating material), the intrinsic stress dominate over the thermal stress. It is important to know whether the thermal stress also plays a role in the residual stress of the system.

In general, the origin of residual stress can be attributed to the epitaxial stress (lattice mismatch between substrate and the film), thermal stress (difference in thermal co-efficient of expansion between substrate and the film), microstructure and the mode of growth (intrinsic stress). The thermal expansion coefficient of AlN film is larger than that of the Si substrate that developed a compressive stress as the sample is cooled down to room temperature from a high temperature since the film thickness is very negligible as compared to the substrate thickness. This thermal stress can be calculated using the relation\cite{Chuen}

\begin{equation}
\sigma_T = \frac{E_f}{(1 - \nu_f)}(\alpha_s - \alpha_f)(T_s - T_r)
\end{equation}

where $\sigma_T$ is thermal stress,$\alpha_s$ and $\alpha_f$ are thermal coefficient of expansion of substrate (Si, $3.6 \times$ 10$^{-6}$ $^\circ$C$^{-1}$) and the AlN film (5.3, $4.2 \times 10$ $^{-6}$ $^\circ$C$^{-1}$ along a-axis and c- axis, respectively) $T_r$ and $T_s$ are room and substrate temperature, $E_f$ and $\nu_f$ are Young's modulus (AlN, 350 GPa) and the Poisson's ratio of AlN film (0.21) \cite{Yim}. The calculated thermal stress is shown in Table~\ref{tab:4}.

\begin{table}[h] 
\centering
\begin{tabular}{l*{1}{c}r}
\hline\hline
$T_s$($^\circ$C) & \vtop{\hbox{\strut $\sigma_T$(a-axis)}\hbox{\strut (GPa)}} & \vtop{\hbox{\strut $\sigma_T$(c-axis)}\hbox{\strut (GPa)}}\\
\hline
35 & -0.007 & -0.002\\
100 & -0.056 & -0.019\\
200 & -0.131 & -0.046\\
300 & -0.207 & -0.073\\
400 & -0.282 & -0.099\\
500 & -0.357 & -0.126\\
600 & -0.433 & -0.152\\
\hline\hline
\end{tabular}
\caption{\label{tab:4}Thermal stress of a-axis and c-axis AlN films at different substrate temperatures.}
\end{table}

Intrinsic or growth stress is a structure and micro-structure sensitive property which is observed in films during deposition and growth under non-equilibrium conditions. In the early stages of film growth, the film consists of smaller crystallites/grains, separated by an atomic-scale distance. When these crystallites/grains coalesce, a tensile stress is generated at the point where neighboring islands impinge on each other \cite{Brian}. As well as, the parameter $T_s/T_m$ is very important in stress-related behavior for different materials. For higher melting point materials, the parameter  $T_s/T_m$ is typically low during deposition. So, under these conditions intrinsic stress can dominate over thermal stress \cite{John,Henry}. Because, when these AlN films deposited at room temperature, there is a little surface diffusion and one would expect the Al or N vacancy concentration to be much larger than at equilibrium. When these excess vacancies are subsequently annihilated at vertical boundary, the associated volume change results in an in-plane change of film dimensions \cite{Song}. So at lower growth temperatures, stress relaxation by diffusion does not occur where intrinsic stress dominates over the thermal stress and a tensile stress develops. 

Also, according to atomic peening model, at high deposition pressure and high TSD (14 cm), the mean free path of the sputtered particle reduces, and there is a particle energy loss due to increase in collision probability between plasma particles which increases the oblique angle particle bombardment. So films contain a tensile stress, due to an atomic self-shadowing effect \cite{John,Henry}. But with the increase in  substrate temperature, the diffusion of defects and vacancies in AlN lattice microstructure causes the decrease in residual stress due to film densification by decreasing the void regions through enhanced atomic rearrangement. Above 400 $^\circ$C, the thermal stress is dominating. So the residual stress changes from tensile to compressive stress \cite{Koch}. It can be reasoned in Fig.\ref{fig:8} that substrate temperature plays a major role on residual stress of sputtered AlN films. Compressive stress at higher temperatures indicates the atomic peening mechanism \cite{Henry}. Such compressive residual stress in coatings usually is expected to improve the quality of the cutting tools and engine cylinder coatings.

\subsection{Nanomechanical properties}
\begin{figure}[h]
\centering
\includegraphics[width=0.60\textwidth]{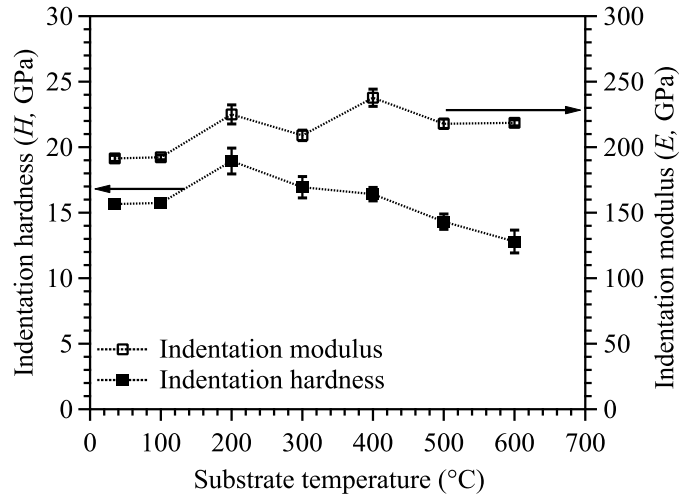}
\caption{Variation in hardness and modulus against substrate temperature.}
\label{fig:9}
\end{figure}

Nanomechanical characterization is essential to understand and solve the problem of controlling residual stress and defect density in the film for design of devices. Nanoindentation studies were performed to estimate the mechanical properties of these films at the peak load of 4000 $\mu$N. The absence of creep and thermal drift was verified by holding the indenter at maximum load for 2 sec (no change in penetration depth). Initially, the hardness $(H)$ increased steadily with substrate temperature until it reached 200 $^\circ$C, where a maximum $H$ of 19 GPa was observed (Fig.\ref{fig:9}). Then, further increase in substrate temperature led only to a constant drop in hardness, down to 12.8 GPa for a substrate temperature 600 $^\circ$C.

In general, orientation of the film, crystallite size and residual stress collectively contributes to the improvement of mechanical properties of the films. Recalling the evolution of the growth structure (TEM (Fig.\ref{fig:5}) and AFM (Fig.\ref{fig:6})); the stress relaxation occurred in larger crystallites (at higher substrate temperatures) can be seen from crystallite size, average grain size and residual stress of these films in Fig.\ref{fig:4}, \ref{fig:7} and \ref{fig:8}, respectively. Generally, the transition from Hall-Petch to inverse Hall-Petch occurs when crystallite size is around 30 nm (supplementary data)\cite{Guisbiers}. On the other hand, nanoindentation modulus $(E)$ of these AlN films is also varied from 191.5 to 237.6 GPa with different substrate temperature. Generally, modulus depends on the nature of inter-atomic bonds in the crystal. At 400 $^\circ$C substrate temperature, AlN film is highly a-axis oriented which contains $B_1$ type of bonds (covalent). So this film has exhibited high elastic nature with higher modulus of 237.6 GPa.

\begin{figure}[h]
\centering
\includegraphics[width=0.60\textwidth]{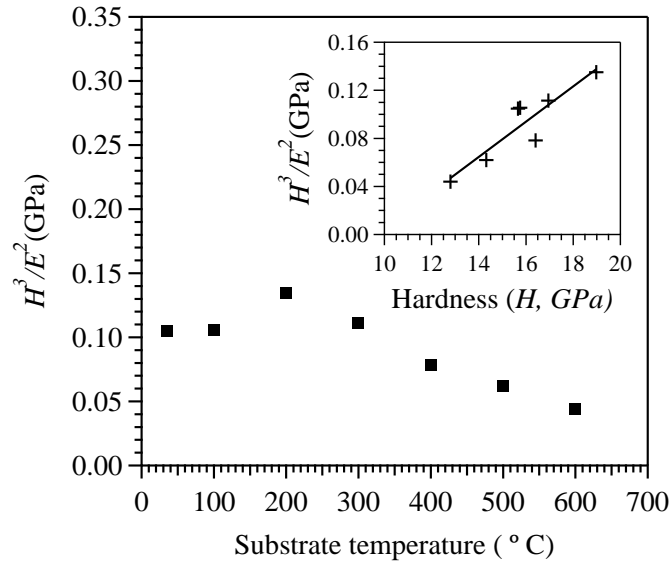}
\caption{Variation in $H^{3}/E^{2}$ against substrate temperature of AlN films. The $H^{3}/E^{2}$ is plotted against indentation hardness in the inset.}
\label{fig:10}
\end{figure}

However, it is possible to produce films with the same $H$ and different values of $E$ i.e. films with the same hardness can exhibit different resistance to plastic deformation. For a protective film such as AlN, it is important to understand whether hardness value is giving true resistance to plastic deformation during contact event. For that the ratio of $H^{3}/E^{2}$ has been widely applied to evaluate the resistance to plastic deformation from the value of $H$ and $E$ \cite{Mayrhofer}. Fig.\ref{fig:10} shows the value of $H^{3}/E^{2}$ of AlN films as a function of substrate temperature. The resistance to plastic deformation has increased from 0.04 to 0.12 GPa. So the parameter $H^{3}/E^{2}$ is also following same profile as the hardness with the substrate temperature. It  is also evident from the inset of Fig.\ref{fig:10} that there is a linear relationship between $H^{3}/E^{2}$ with hardness  which agrees with Mayrhofer \emph{et al}. 

\section{Conclusion}
DC reactive magnetron sputtering has been used to synthesize AlN thin films on Si (100) substrate at different substrate temperatures varying from 35 to 600 $^\circ$C. All these films crystallized with wurtzite structure and the substrate temperature influenced the orientation of these films. The crystallinity increased with temperature and preferred orientation (a-axis) was observed at 400~$^\circ$C. The cross section and SAED pattern of AlN films showed a high degree of alignment (columnar structure) as well as the orientation. The surface morphology, RMS roughness and average grain size are also strongly influenced by the substrate temperature. However the residual stress measurement by $sin^{2}\psi$ technique showed tensile stress up to 300 $^\circ$C and then changed to compressive from 400 $^\circ$C to 600 $^\circ$C. Nano-indentation hardness studies on these films revealed that the indentation hardness $(H)$ varied between 12.8 to 19 GPa, where as at 400~$^\circ$C substrate temperature AlN film exhibits a relatively high elastic nature ($E$ = 237.6 GPa).

\section{Acknowledgement}
The authors would like to thank UGC-DAE Consortium for Scientific Research, Kalpakkam, India for the GIXRD facility and Dr. Chintan M. Bhatt (Hysitron Nanotechnology, India) for nanoindentation.


\section{References}
\begin{thebibliography}{99}

\bibitem{Hadis}Hadis Morkoc, Handbook of Nitride Semiconductors and Devices.Vol.1 : Materials Properties, Physics and Growth (WILEY-VCH, Weinheim, 2008) p. 23.
\bibitem{Pan}T. S. Pan, Y. Zhang, J. Huang, B. Zeng, D. H. Hong, S. L. Wang, H. Z. Zeng, M. Gao, W. Huang, and Y. Lin 2012 J. Appl. Phys. 112 044905.
\bibitem{Barkad}H. A. Barkad, A. Soltani, M. Mattalah, J.C. Gerbedoen, M. Rousseau, J.C. De Jaeger, A. BenMoussa, V. Mortet, K. Haenen, B. Benbakhti, M. Moreau, R. Dupuis, and A. Ougazzaden 2010 J. Phys. D: Appl. Phys. 43 465104.
\bibitem{Dimitrova}V. Dimitrova, D. Manova, E. Valcheva 1999 Mater. Sci. Eng. B 68 1.
\bibitem{Matt}Matt D. Brubaker, Igor Levin, Albert V. Davydov, Devin M. Rourke, Norman A. Sanford, Victor M. Bright, and Kris A. Bertness 2011 J. Appl. Phys. 110 053506.
\bibitem{Taniyasu} Yoshitaka Taniyasua and Makoto Kasu, 2010 Appl. Phys. Lett. 96 221110. 
\bibitem{Pei}PeiTsen Wu, Mitsuru Funato \& Yoichi Kawakami 2015 Sci. Rep. 5, 17405.
\bibitem{Shibo}Shibo Yang, Reina Miyagawa, Hideto Miyake, Kazumasa Hiramatsu and Hiroshi Harima 2011 Appl. Phys. Exp. 4 031001.
\bibitem{Bryan}I. Bryan, A. Rice, L. Hussey, Z. Bryan, M. Bobea, S. Mita, J. Xie, R. Kirste, R. Collazo, and Z. Sitar 2013 Appl. Phys. Lett. 102 061602.
\bibitem{Georgieva}A. Kakanakova-Georgieva, G. K. Gueorguiev, R. Yakimova and E. Janzen 2004 J. Appl. Phys. 96 5293.
\bibitem{Kuyoma}T. Kuyoma, M. Sugawara, Y. Uchiyama, J. F. Kaeding, R. Sharma, T. Onuma, S. Nakamura, and S. F. Chichibu 2006 Phys. Status Solidi A 203 1603.
\bibitem{Okamoto}K. Okamoto, S. Inoue, T. Nakano, T.-W.Kim, M. Oshima, and H. Fujioka 2008 Thin Solid Films 516 4809.
\bibitem{Liu}H.Y. Liu, G.S. Tang, F. Zeng, F. Pan 2013 J. Cry. Growth 363 80.
\bibitem{Ping}Xu-Ping Kuang, Hua-Yu Zhang, Gui-Gen Wang, Lin Cui, Can Zhu, Lei Jin, Rui Sun, Jie-Cai Han 2012 Superlattices and Microstructures 52 931.
\bibitem{Mirpuri}C. Mirpuri, S. Xu, J. D. Long, and K. Ostrikov 2007 J. Appl. Phys. 101 024312.
\bibitem{Xu}X.H. Xu, H.S. Wu, C.J. Zhang, Z.H. Jin 2001 Thin Solid Films 388 62.
\bibitem{Feby}Feby Jose, R. Ramaseshan, S. Tripura Sundari, S. Dash, A. K. Tyagi, M. S. R. N. Kiran, and U. Ramamurty 2012 Appl. Phys. Lett. 101 254102.
\bibitem{Soh}Martin T. K. Soh, A. C. Fischer-Cripps, N. Savvides, C. A. Musca and L. Faraone 2006 J. Appl. Phys. 100 024310.
\bibitem{Jose}Feby Jose, R. Ramaseshan, S. Dash, S. Bera, A. K. Tyagi, and Baldev Raj 2010 J. Phys. D: Appl. Phys. 43 075304.
\bibitem{Kuan}Kuan-Hsun Chiu, Jiann-Heng Chen, Hong-Ren Chen, Ruey-Shing Huang 2007 Thin Solid Films 515 4819.
\bibitem{Medjani} F. Medjani, R. Sanjines, G. Allidi and A. Karimi 2006 Thin Solid Films 515 260.
\bibitem{Mahieu}S. Mahieu, P. Ghekiere, D. Depla, R. De Gryse 2006 Thin Solid Films 515 1229.
\bibitem{Polaki}ShyamalRao Polaki, Rajagopalan Ramaseshan, Feby Jose, S. Dash, A. K. Tyagi, Nukala Ravi 2013 Int. J. Appl. Ceram. Technol. 10 45.
\bibitem{Wang}J. Wang, Q. Zhang, G. F. Yang, C. J. Yao, Y. J. Li, R. Sun, J. L. Zhao \& S. M. Gao  2016 J Mater Sci: Mater Electron. 27, 3026.
\bibitem{Knisely}Katherine Knisely \& Karl Grosh 2014 J. Vac. Sci. Technol. A 32, 051504.
\bibitem{Dubois}Marc-Alexandre Dubois \& Paul Muralt 2001 J. Appl. Phys. 89, 6389.
\bibitem{Si}Si-Hyung Lee, Ki Hyun Yoon, Deok-Soo Cheonga \& Jeon-Kook Lee 2003 Thin Solid Films 435, 193.
\bibitem{Chuen}Chuen-Lin Tien, Tsai-Wei Lin, Kuo-Chang Yu, Tsung-Yo Tsai and Hsi-Fu Shih 2014 IEEE TRANSACTIONS ON MAGNETICS, 50(7) 2005304.
\bibitem{Yim}W. M. Yim and R. J. Paff 1974 J. Appl. Phys. 45 1456. 
\bibitem{Brian}Brian W. Sheldon, K. H. A. Lau, and Ashok Rajamani 2001 J. Appl. Phys. 90 5097.
\bibitem{John}John A. Thornton and D. W. Hoffman 1989 Thin Solid Films, 171 5.
\bibitem{Henry} Henry Windischmann 1992 Crit. Rev. Solid State Mater. Sci. 17, 547.
\bibitem{Song}J.Y. Song, Jin Yu 2002 Thin Solid Films 415 167.
\bibitem{Koch} R. Koch 2010 Surf. Coat. Technol. 204, 1973.
\bibitem{Guisbiers} G Guisbiers and L Buchaillot 2008 J. Phys. D: Appl. Phys. 41, 172001.
\bibitem{Mayrhofer}P.H. Mayrhofer, C. Mitterer \& J. Musil 2003 Surf.Coat.Technol. 174, 725.

\end{thebibliography}
\end{document}